\documentclass{article}

\usepackage{arxiv}

\usepackage[utf8]{inputenc} 
\usepackage[T1]{fontenc}    
\usepackage{hyperref}       
\usepackage{url}            
\usepackage{booktabs}       
\usepackage{amsfonts}       
\usepackage{nicefrac}       
\usepackage{microtype}      
\usepackage{lipsum}

\usepackage{float}
\usepackage{morefloats}
\usepackage{color}
\usepackage{multirow}
\usepackage{latexsym}
\usepackage{amsmath,amssymb,amsthm,graphicx}
\usepackage{dsfont}
\usepackage{enumitem}
\usepackage{hyperref}
\usepackage{arydshln}
\usepackage{lscape}
\usepackage{adjustbox}

\usepackage{todonotes}
\DeclareMathAlphabet{\mathscr}{OT1}{pzc}{m}{it}

\newtheoremstyle{def}
{9pt}
{9pt}
{}
{}
{\bfseries}
{.}
{ }
{}
\theoremstyle{def}

\renewcommand{\footnoterule}{%
	\kern -3.5pt
	\hrule width \textwidth height 1pt
	\kern 3.5pt
}

\makeatletter
\def\blfootnote{\xdef\@thefnmark{}\@footnotetext}
\makeatother

\title{A critical review of existing and new population stability testing procedures in credit risk scoring}

\author{J. du Pisanie\\
School of Mathematical and Statistical Sciences,\\ North-West University,\\ South Africa. \\
\href{mailto:dupisanie@gmail.com}{johan.dupisanie@monoclesolutions.nl}\\
\And J. S. Allison\\
School of Mathematical and Statistical Sciences,\\ North-West University,\\ South Africa.\\
\href{mailto:james.allison@nwu.ac.za}{james.allison@nwu.ac.za}
\And C. J. Budde\\
Department of Mathematical and Applied Mathematics,\\ University of the Free State,\\ South Africa.\\
\href{mailto:buddecj@ufs.ac.za}{buddecj@ufs.ac.za}
\And I. J. H. Visagie\\
School of Mathematical and Statistical Sciences,\\ North-West University,\\ South Africa.\\
\href{mailto:jaco.visagie@nwu.ac.za}{jaco.visagie@nwu.ac.za}\\
}

\begin{document}

\date{\today}
\maketitle

\begin{abstract}
    Credit scorecards are models used for the modelling of the probability of default of clients. The decision to extend credit to an applicant, as well as the price of the credit, is often based on these models. In order to ensure that scorecards remain accurate over time, the hypothesis of population stability is tested periodically; that is, the hypothesis that the distributions of the attributes of clients at the time when the scorecard was developed is still representative of these distributions at review is tested. A number of measures of population stability are used in practice, with several being proposed in the recent literature. This paper provides a critical review of several testing procedures for the mentioned hypothesis. The widely used population stability index is discussed alongside two recently proposed techniques. Additionally, the use of classical goodness-of-fit techniques is considered and the problems associated with large samples are investigated. In addition to the existing testing procedures, we propose two new techniques which can be used to test population stability. The first is based on the calculation of effect sizes which does not suffer the same problems as classical goodness-of-fit techniques when faced with large samples. The second proposed procedure is the so-called overlapping statistic. We argue that this simple measure can be useful due to its intuitive interpretation. In order to demonstrate the use of the various measures, as well as to highlight their strengths and weaknesses, several numerical examples are included.
\vspace{0.5cm}



\end{abstract}

\section{Introduction and motivation}
\label{Intro}

In the banking industry, models are widely used to quantify, among other things, various types of risk associated with a customer or applicant. Some of the best known of these models include those for the probability of default (PD), the loss given default (LGD) and the exposure at default (EAD). Note that the definition of default is set by the modelling team of the bank, but default usually refers to a customer that has missed 3 payments in a 12 month cycle. The variations in this definitions are beyond the scope of the current paper and we use the mentioned definition for the remainder of the discussion.

PD, LGD and EAD models can be used separately or combined to assess risk and determine the level of caution needed when dealing with a given customer. When used together, the PD model is often considered to be the most influential of the three models. The reason being the PD associated with a customer is often more variable than either the LGD or EAD, making the PD the main risk driver for the individual customer. Depending on the financial product under consideration, PDs between 1\% and 10\% are commonplace and a small change in the average PD can potentially have a substantial impact on exposure to risk. One of the main models used for PDs is the logistic regression model, commonly known as a scorecard.
This model provides a predicted PD given various attributes of a customer (we refer to the characteristics of the customer included in the model as attributes throughout). Scorecards are used extensively in banks and other financial institutions. As highlighted in \cite{DUPISANIEVISAGIE2020}, there are three main areas in which scorecards are used:
\begin{enumerate}
\item \emph{Credit risk scorecards} – These scorecards are used to estimate the PD of a new applicant who wishes to take up some financial product as well as to estimate the PD of existing customers when applying to take up additional credit. Additionally, these models are instrumental when setting the credit strategy as well as in the calculation of loan loss provisions.
\item \emph{Collection scorecards} – These scorecards are used to determine the probability of repayment for customers with one or more accounts in arrears. The outcomes of the scorecards are used to determine where collections efforts will be most effective.
\item \emph{Marketing scorecards} – These scorecards are used to estimate the probability that a given potential customer will take up an offer. The application of these scorecards enables the business to focus marketing efforts, usually through outbound calling or engagements, on the customers most likely to take up a specific offer.
\end{enumerate}

The average accuracy of a scorecard can be measured by comparing the predicted and observed outcomes. However, this comparison can only be performed once the outcome period is complete. By way of an example, say a business were to on-board (approve and take on) 100 customers with an average probability of default of 5\% during the next 12 months. In this case, if the scorecard is accurate, we expect that 5 customers will default during the outcome period of 12 months. It is possible to review the 100 customers after 12 months and compare the number of defaults to the predicted value of 5. If the default number is found to differ substantially from 5, it serves as an indication that the model does not accurately predict default. One possible explanation for this phenomenon is an inherent change in the level of risk associated with customers having specified attributes. A second possible explanation is a material change in the structure of the observed characteristics of the population of customers; i.e., the distribution of the attributes may have changed to such an extent that the model is no longer representative of the population under consideration. Our interest centers on the second of these potential reasons, meaning that we are interested in identifying a substantial change in the distribution of the attributes of customers. If this distribution changes over time, there is reason to doubt the predictions made by the model at a later stage. In order to ensure that the model can still be used, practitioners periodically test the hypothesis that the distribution of the attributes at present is the same as the the distribution observed when the model was fitted.

Before proceeding to a detailed account of the methods used to identify changes in the distribution of the attributes, a distinction should be made between a change in the distribution of an attribute and instability in the measurement of an attribute. A change in the distribution of an attribute indicates that the type of customer being scored through the model has changed. For example, the data used in order to develop the scorecard (that is, to fit the logistic regression model) may have contained a high percentage of male applicants whereas the current composition of the customers contains a higher percentage of female applicants. This is the type of change that we wish to identify using some statistically sound procedure. However, the measurement of a characteristic can also become unstable as a result of some external influence which does not truly have any effect on the accuracy of the model. As a concrete example, consider a variable that scores customers with more than two credit enquires in the last two months as high risk. If a new credit provider enters the market and scores the population for marketing purposes, then every potential customer will have an additional enquiry to their name and will subsequently be perceived as being a higher risk, even though the risk associated with the potential customer has remained unchanged. The latter type of changes are not considered in the remainder of the paper.


As was mentioned above, the accuracy of a scorecard can easily be measured following the outcome period. However, the bank will typically be interested in measuring the accuracy of the scorecard on a continuous basis in an attempt to ensure that the estimated PDs are as accurate as possible. To achieve this, banks make use of population stability metrics (PSMs). These are essential tools that enable the business to validate scorecards, to some extent, before the actual outcome becomes available. These metrics aim to determine whether the distribution of the attributes of the current customers
is sufficiently similar to that of the customers used to develop the scorecard to ensure that the model remains representative. This paper provides an overview of the use of PSMs in relation to scorecards.


\newpage

A few general remarks regarding the use of a PSMs should be made at the outset;
\begin{itemize}
\item PSMs will not identify factors lying outside of the scorecard. For instance, the population and outcome distributions for a marketing scorecard might be stable, yet the agent making the sales call could have an impact on the outcome.
\item The economic impacts on a collections scorecards may not be immediately apparent, as they could be delayed. As a result, a collections scorecard may remain stable even if the population it represents is experiencing economic hardships, such as higher inflation or interest rates, which could decrease their ability to pay. 
\item PSMs cannot distinguish between a population shift and an unstable population characteristic. Once a PSM indicates that the distribution of the underlying data has changed, it will be necessary for the business to understand the reason for the shift via further investigation as to the reason for the change.
\item In this paper, we do not consider the impact of a detected change on the outcome. The PSMs considered in this paper aim to detect the presence of a change in the distribution of the attribute and not to examine the effect of this change on the estimated PDs.
\end{itemize}

Despite the caveats made above, PSMs can provide an early indication to the business that the distribution underlying the population is changing and that the calculated PDs might not be in-line with what actually transpires with the customers on book.
This paper reviews alternative measures of population stability and provides insights into the considerations for choosing a PSM. Additionally, we consider two new approaches which may be useful for measuring population stability. The first is based on calculated effect sizes, while the second is a measure referred to as the overlapping statistic. This statistic has been used in the context of psychology and we demonstrate that it can be used as an intuitive measure of population stability. 

The remainder of this paper is structured as follows. Section 2 provides a brief overview of the scorecard building process while Section 3 details the statistical considerations relating to PSMs. Special attention is paid to the problems associated with large sample sizes and the advantages and disadvantages of each metric is outlined. Section 4 provides numerical examples illustrating the use of the PSMs in practice. Conclusions and recommendations are provided in Section 5.

\section{A review of the scorecard building process}

This section contains an overview of the process used in order to build a scorecard. The procedure used is described in more detail in \cite{SIDDIQI2016}. Below we briefly touch on the major topics required for the understanding of scorecard building and monitoring.

\subsection{Key concepts}

Scorecards are tools used to model the PDs of customers. For a given customer, the attributes are each assigned a score. These individual scores are aggregated into what is referred to as a credit score, whence the name scorecard.
The model translates the credit score into an estimated PD. This PD is usually what the credit provider bases credit decisions on.

\subsection{Scorecard building}


The scorecard building process typically consists of the following steps:
\begin{enumerate}
    \item \textbf{Obtaining data} - A data set is identified for the building of the scorecard. The data set should contain the attribute values at the time of the original decision; for example when the customer applied for a loan. It should also contain the outcome attribute of the customer to indicate whether the customer defaulted.
    \item \textbf{Consider segmentation} - If enough data is available, and if the customers can be divided into distinct segments, this should be considered. An analysis is needed to understand the optimal segments. Segments should be chosen such that their underlying properties of the attributes are vastly different across segments. Once the segmentation is decided upon, a separate scorecard can be built for each segment.
    \item \textbf{Grouping attribute values} - The range for each attribute should be inspected for grouping. The process of creating groups serves a dual purpose. First, it creates stability in the scorecard. By way of an example, consider a customer's income. It is often practical to divide clients into a number of groups based on their income since it is reasonable to expect the modelled PD will change gradually with income. It therefore makes sense to group the attributes into buckets where the behaviour towards the PD is the same. In order to determine a reasonable grouping associated with an attribute, the information value (discussed below) is used in conjunction with practical considerations.
    \item \textbf{Calculate the weights of evidence (WOEs)} - The purpose of calculating the WOEs, is that it standardizes the range of each attribute. When creating the statistical model, the magnitude of the estimated coefficients in the model can be compared to one another. Consider a discrete attribute with $k$ levels. Let $D$ be the number of defaults in the data set, let $D_j$ be the number of defaults associated with the $j^{\textrm{th}}$ level of this attribute and let $n_j$ be the total number of observations associated with the $j^{\textrm{th}}$ level of this attribute. The WOE is then calculated as follows:
\begin{equation*}
    {WOE_j}=\textrm{log}\left(\frac{D(n_j-D_j)}{D_j(n-D)}\right).
\end{equation*}
    \item \textbf{Select the attributes used in the scorecard} - Depending on the number of available attributes, a process of elimination can be employed to remove some attributes. The considerations include the availability of the attribute in future; an attribute that might not be available in future should be removed. Furthermore, legal restrictions relating to the use of each attribute should be considered. Attributes such as race or gender are usually not allowed in a credit scorecard as it can create bias in the decision-making process.  Finally, the strength of the association between the attribute and the PD should be considered; only attributes which have a practically significant relationship with PD should be included in the model. The predicting ability of an attribute in the scorecard can be measured by the Information Value (IV), see \cite{SIDDIQI2006}:
\begin{equation*}
    \textrm{IV}=\sum_{j=1}^k \left(\frac{n_j-D_j}{n-D} - \frac{D_j}{D}\right) \textrm{log}\left(\frac{D(n_j-D_j)}{D_j(n-D)}\right).
\end{equation*}
    Siddiqi \cite{SIDDIQI2006} proposes the following rule of thumb for the interpretation of the IV:
    \begin{itemize}
    \item If the IV is less than 0.02, then the attributed is considered unpredictive.
    \item An IV of between 0.02 and 0.1 is a associated with a weak predictive ability.
    \item If the IV is between 0.1 and 0.3, then the predictive ability of the attribute is classified as medium.
    \item A strong predictive ability is associated with an IV of 0.3 or more.
    \end{itemize}
    If there exists high correlations between two variables, it is recommended that only one be included in the final scorecard so as to avoid problems associated with multicollinearity.
    \item \textbf{Perform a logistic regression} - The model for the scorecard is built by fitting a logistic regression to the WOEs of all the selected attributes.
\end{enumerate}

\subsection{Scorecard monitoring on underlying and grouped data}

The monitoring of the scorecard now presents the opportunity to consider both the underlying, possibly continuous, data as well as the grouped data. It is advisable to investigate the distribution of both when testing for population stability. The testing of the grouped data provides insights into the effect that the underlying stability of the population has on the scorecard and outcome. This should be the main indicator to show that the distribution of the attributes has remained stable or has become unstable. In the latter event, a decision to rebuild or decommission a scorecard based on data stability may be taken.

It is also advisable to test the underlying, ungrouped data for stability. Such a test could serve as an early indicator that, even though the model groupings remain stable, the underlying data is shifting away from the original distribution. In this case, further investigation is required in order to understand if such a trend is likely to continue and, if so, what the impact on the predictions made using the scorecard would be.

\subsection{Impact of population instability on the overall PD outcome}

We assume throughout that predicted PDs are calculated at account level. This is generally the case in practice, especially in developed credit markets where data availability is not a problem. When calculating these PDs, the average PD of the entire population is often calculated in order to quantify the inherent level of risk associated with the population as a whole.

When certain attributes in a population becomes unstable, it is possible that the effect of these attributes on the overall PD offset one another to a certain degree. Consider two attributes becoming unstable, however the one attribute moves applicants to a lower PD score and the other attribute moves applicants to a higher PD score. Depending on the magnitude of the two shifts, the impacts could offset one another, meaning that the overall level of risk may remain unchanged. An analyst only considering the average PD of the population to decide whether the population remains stable will not identify this move correctly and will falsely conclude that the population, and all its attributes, are stable.

\section{A review of the literature on population stability metrics}\label{PSMs}

This section contains a review of the available literature on PSMs which can be employed to test the hypothesis of population stability.
Before going into the details of these PSMs, we provide a general discussion on the hypothesis of population stability. This discussion includes comments on the nature of the hypothesis to be tested as well as the methods required in order to arrive at interpretations of the calculated values of these measures. We discuss two different approaches to these interpretations; one aimed at determining statistical significance and the other based on rules of thumb. In the case of the former, we require the calculation of critical values. Below, we include an algorithm which can be used in order to approximate the critical values associated with the various PSMs considered.

After the general discussion regarding the hypothesis to be tested, we turn our attention to the various PSMs available in the literature. These include the population stability index (PSI) which has become something of an industry standard. Two recently proposed PSMs are also considered; these measures are proposed in Taplin and Hunt \cite{TAPLINHUNT2019} as well as Du Pisanie and Visagie \cite{DUPISANIEVISAGIE2020}. Thereafter we turn our attention to classical statistical techniques; we consider the Kolmogorov-Smirnov goodness-of-fit test as well as a measure based on effect sizes. This section concludes with a discussion about the so-called overlapping statistic which measures the similarity of distributions. This measure has not been used to test population stability in credit scoring to date. However, we argue that this useful and intuitive measure can provide analysts with valuable insight into the questions surrounding population stability.

\subsection{The hypothesis to be tested}\label{HTBT}

As mentioned above, the distributional properties of the attributes and outcome are monitored over time. Some of these attributes are continuous (such as income) while others are discrete (such as the number of credit cards used by the customer). It is explained above that continuous attributes are typically divided into a number of groups and are then treated as discrete in the scorecard.
As a result, our main interest centers around testing the hypothesis that a discrete random variable follows a specified distribution.

In order to proceed, we introduce notation similar to that used in \cite{DUPISANIEVISAGIE2020}. Consider some attribute with $k$ levels; let $\mathbf{q}=(q_1,\dots,q_k)$ denote the proportions associated with each of the values that this attribute can assume at the time of development of the scorecard. Denote by $\mathbf{p}=(p_1,\dots,p_k)$ and $\mathbf{P}=(P_1,\dots,P_k)$ the population and sample proportions associated with the $k$ levels of the attribute, respectively, at the time at which population stability is to be tested. The hypothesis of interest is
\begin{equation}\label{eqn:H0}
    H_0: p_j=q_j, \forall j \in \{1,\dots,k\},
\end{equation}
which is to be tested against general alternatives.

It is common practice in the building of scorecards to group the predicted outcome into so-called PD groups. These groups are often selected to be similar in size; i.e. the top $x$\% of the PDs are grouped together, then the next $x$\% and so on. The grouping of the PDs can be used in the strategy of the product in questions, such as assigning certain limits or giving benefits to customers in higher PD groups. The PSMs being discussed in this paper can be used to test the stability of the PD groups in the same way that any attribute in the scorecard is tested.

Since $\mathbf{p}$ cannot be observed directly, the hypothesis in (\ref{eqn:H0}) is tested based on $\mathbf{P}$ (as well as $\mathbf{q}$). As a result, the majority of the PSMs considered constitute discrepancy measures between $\mathbf{q}$ and $\mathbf{P}$.
That is, given a specific attribute, PSMs generally aim to quantify the differences between the distribution at the time that the scorecard is developed and the time at which the population stability is tested. To formalise this, a PSM is typically a discrepancy measure, or a measure of dissimilarity between $\mathbf{q}$ and $\mathbf{P}$. Let $\Delta(\mathbf{q},\mathbf{P})$ denote some PSM based on the specified value of $\mathbf{q}$ and the observed value of $\mathbf{P}$.

In general, there are two approaches employed when interpreting a realised value of $\Delta(\mathbf{q},\mathbf{P})$. The first is to interpret the value relative to a specified rule of thumb, while the second is to approximate a critical value under the hypothesis of population stability. Both of these approaches are widely criticized in the relevant literature but each has some merit. While rules of thumb are difficult to justify from a statistical perspective, the large sample sizes often encountered in practice cast doubt as to the use of calculated $p$-values. However, the calculation of critical values and $p$-values are useful in order to study the behaviour of $\Delta(\mathbf{q},\mathbf{P})$ in the case where the hypothesis in \eqref{eqn:H0} holds. Below, we consider the required calculations to arrive at critical values and $p$-values.

Let $m$ denote the number of customers in the current sample to be tested for population stability. When testing the population stability of some attribute, $\mathbf{q}$ is a fixed constant vector. Under the assumption that \eqref{eqn:H0} is true, we have that $\mathbf{p}=\mathbf{q}$. In this case, $m\mathbf{P}$ follows a multinomial distribution with parameters $m$ and $\mathbf{q}$. As a result, we can easily simulate realisations of $\mathbf{P}$ under the null hypothesis. Note that no observed value of $\mathbf{P}$ is required in order to perform this simulation, all that is required is that $\mathbf{q}$ and the sample size $m$ be known. The critical value of $\Delta$ can now be approximated using the following algorithm.
\begin{enumerate}
\item Given $\mathbf{q}$ and $m$, simulate a realisation, say $\mathbf{M}^*$, from a multinomial distribution with parameters $m$ and $\mathbf{q}$. Calculate $\mathbf{P}^*=\mathbf{M}^*/m$. 
\item Calculate $\Delta^*=\Delta(\mathbf{q},\mathbf{P}^*)$.
\item Repeat the first two steps above $b$ times in order to obtain $b$ realisations of $\Delta$ under the null hypothesis. Denote the results by $\Delta_1^*,\ldots,\Delta^*_{b}$.
\item Order the observed values to obtain $\Delta^*_{(1)} \leq \dots \leq \Delta^*_{(b)}$.
\item The approximate critical value at a $100\alpha \%$ significance level is $\Delta_{(\beta)}$, where $\beta=\lfloor b(1-\alpha) \rfloor$ and $\lfloor \cdot \rfloor$ is the floor function.
\end{enumerate}
Given an observed value of $\mathbf{P}$, the calculation of the p-value associated with $\Delta(\mathbf{q},\mathbf{P})$ is a simple matter. This $p$-value is
\begin{equation*}
    \frac{1}{b} \sum_{j=1}^b \textrm{I}\left(\Delta(\mathbf{q},\mathbf{P}) \leq \Delta^*_{j}\right),
\end{equation*}
where $\textrm{I}$ denotes the indicator function.

Note that the algorithm for the calculation of empirical critical and $p$-values above can be used for each of the PSMs considered below with the exception of the overlapping measure. However, altering the formulas above for use with this measure is straightforward. Let $\xi(\mathbf{q},\mathbf{P})$ be the overlapping statistic. Upon setting $\Delta(\mathbf{q},\mathbf{P})=1-\xi(\mathbf{q},\mathbf{P})$, the calculations above can be used to obtain critical values and $p$-values for $\Delta$ and, by implication, for $\xi$.



\subsection{The population stability index (PSI)} \label{psi}


The PSI is the most widely used PSM and can be found in several texts, including \cite{Lew1994} and \cite{SIDDIQI2006}. Using the notation defined above, the PSI is defined to be
\begin{equation}
    \Psi(\mathbf{q},\mathbf{P}) = \sum_{j=1}^{k} (P_j - q_j)\log \left(\frac{P_j}{q_j}\right).
\end{equation}
Lewis \cite{Lew1994} proposes the following interpretation of the realised value of the PSI:
\begin{enumerate}
    \item $0 \leq \Psi(\mathbf{q},\mathbf{P}) < 0.1$ indicates that the population shows no substantial change.
    \item $0.1 \leq \Psi(\mathbf{q},\mathbf{P}) < 0.25$ shows a small change in the population.
    \item $\Psi(\mathbf{q},\mathbf{P}) \geq 0.25$ indicates a substantial change in the population.
\end{enumerate}

The PSI is considered and discussed in several recent papers; see for example \cite{YN2019} as well as \cite{DUPISANIEVISAGIE2020}. It should be noted that the mentioned papers are critical of the PSI and suggest that, although this measure is useful for testing population stability, it should be used with care.

Yurdakul and Naranjo \cite{YN2019} is concerned with the statistical properties of the PSI. In addition to showing that, a properly rescaled version of the PSI, asymptotically follows a $\chi$-square distribution, the authors argue for the use of asymptotic critical values for the interpretation of the PSI rather than the use of rule of thumb described above. It is shown that the asymptotic critical values depend on both the sample size and the number of groups associated with the attribute in question.

Du Pisanie and Visagie \cite{DUPISANIEVISAGIE2020} is critical of the PSI since the interpretation of the calculated numerical value of $\Psi$ does not take the sample size or number of levels associated with the relevant attribute into account. The mentioned paper provides numerical examples illustrating that the critical value of the PSI is a function of both the sample size and number of levels; the critical value of the PSI decreases with sample size and increases with the number of groups associated with the attribute (at least in the case where the proportions associated with the various groups are equal). In the extreme case where one of levels present in either the development or review populations is not present in the other, i.e. when a level of the attribute is observed in one population but not the other, the calculation of the $\Psi$ breaks down and this measure is calculated to be infinite.

\subsection{The measure of Taplin and Hunt \cite{TAPLINHUNT2019}}


An alternative approach to scorecard stability testing is provided in Taplin and Hunt \cite{TAPLINHUNT2019}. The authors propose a metric called the population accuracy index (PAI) which measures the change in the variance of the estimated mean response since development. The PAI is defined to be the ratio between two quantities; the numerator is the average variance of the estimated mean response at the time that the scorecard is reviewed, while the denominator is the average variance of the estimated mean response at the development of the scorecard. In the current context, the response is the calculated PD.

Let $\mathbf{X}=(X_1,\dots,X_n)$ and $\mathbf{Y}=(Y_1,\dots,Y_m)$ denote the observed values of the attribute under consideration at development and review, respectively. In the case where a simple linear regression is used in order to model the PD, the PAI can be shown to be
\begin{equation}
        \pi(\mathbf{X},\mathbf{Y}) = \frac{1}{2} \left(1+\frac{\frac{1}{m}\sum_{j=1}^m(Y_j-\bar{X} )^2 }{\frac{1}{n}\sum_{j=1}^n(X_j-\bar{X} )^2}\right), \label{THF}
\end{equation}
where $\bar{X}$ and $\bar{Y}$ denote the sample averages of $X_1,\dots,X_n$ and $Y_1,\dots,Y_m$, respectively. If a multiple linear regression model is employed, the PAI admits a closed form expression once more; the interested reader is referred to \cite{TAPLINHUNT2019} for more details. Smaller values of the PAI is deemed more desirable and indicate that no substantial change in the distribution has occurred. To be specific, the recommended interpretation of the PAI is as follows:
\begin{enumerate}
    \item $0 \leq \pi(\mathbf{X},\mathbf{Y}) < 1.1$ indicates that the distribution of the attribute shows no substantial change.
    \item $1.1 \leq \pi(\mathbf{X},\mathbf{Y}) < 1.5$ shows a small change in the distribution of the attribute.
    \item $\pi(\mathbf{X},\mathbf{Y}) \geq 1.5$ indicates a substantial change in the distribution of the attribute.
\end{enumerate}

While there is certainly merit in considering the accuracy with which the PD can be predicted or calculated, there are certain aspects of the PAI that makes this metric difficult to implement in practice.
First, the PAI is proposed for use with scorecards and, although the measure can be used in the context of a logistic regression model, no closed form expression is available for the PAI in this case. In fact, in order to arrive at the variance of individual calculated PDs in this case, a parametric bootstrap procedure is required. This complexity may prove problematic, especially given the high level of importance afforded the interpretation of PSMs in practice.
Second, the PAI implicitly assumes that a smaller average variance is preferable to a larger variance. This is an arbitrary assumption as both a smaller or larger variance could indicate instability in the underlying population.

To illustrate a possible pitfall with this measure, consider the following toy example. Let the observed values of the attributes at development and review be $\textbf{X}=(-2,1,1)$ and $\textbf{Y}=(1.1,1.2,1.3)$, respectively. In this example there is no overlap between the values of $\textbf{X}$ and $\textbf{Y}$, indicating that the population is unstable in the most extreme manner imaginable. However, consider the calculated PAI using (\ref{THF}):
\begin{eqnarray*}
\pi(\mathbf{X},\mathbf{Y}) = \frac{1}{2} \left(1+\frac{\frac{1}{m}\sum_{j=1}^m(X_j-\bar X )^2}{\frac{1}{n}\sum_{j=1}^n(X_j-\bar X )^2}\right) = \frac{1}{2} \left(1+\frac{1.1^2 + 1.2^2+ 1.3^2}{2^2+1^2+1^2}\right) = 0.862.
\end{eqnarray*}
According to the recommendations made in \cite{TAPLINHUNT2019}, the calculated PAI value should be considered to indicate no substantial change in the distribution of the attribute considered.

Although the PAI has merit as a measure of population stability, we would recommend that this measure not be considered in isolation.




\subsection{The measure of Du Pisanie \& Visagie \cite{DUPISANIEVISAGIE2020}}

Du Pisanie and Visagie \cite{DUPISANIEVISAGIE2020} argues that the hypothesis of population stability, as stated in \eqref{eqn:H0}, is overly strict. That is, the classical requirement for population stability is that the distribution of the attributes remains exactly the same over time. Given the large sample sizes often encountered in practice, it is possible to detect minuscule changes in the distribution over time using standard statistical methods. The mentioned paper argues that, in practice, the more interesting question is whether or not the observed change in the distribution is substantial enough in order to negatively affect the predictions made by the scorecard.

The authors propose an alternative PSM which aims to test whether or not the observed change in the distribution exceeds some user-defined threshold. Using the same notation as above, the proposed PSM is
\begin{equation}
    T(\mathbf{q},\mathbf{P}) = \max_{1\leq j\leq k}\frac{\lvert P_j - q_j \rvert}{q_j}. \label{DVF}
\end{equation}
The realised value of $T$ is compared to a user-defined parameter, say $\delta$. If $T(\mathbf{q},\mathbf{P})$ exceeds $\delta$, then a substantial change in the distribution is deemed to have occurred. The value of $\delta$ is chosen in line with the relative importance of the scorecard. The greater the exposure to loss if the scorecard is not accurate, the smaller the value of $\delta$ should be chosen. An advantage of this PSM is that the user-defined value of $\delta$ provides the business with an opportunity to set the sensitivity of the test to various levels for different scorecards.

Population stability techniques are typically employed to determine if a scorecard is still accurate in its predictions. In the case of credit risk scorecards, there are often attributes which constitute some minimum requirement for credit to be approved. As a result, there will be one or more levels of these attributes for which the accuracy of the predicted PD will be only of academic interest as it will not influence the profitability of the bank. Therefore, \cite{DUPISANIEVISAGIE2020} argues that these classes of the attribute in question should be removed from the analysis. To be precise, let $k^*$ denote the number of groups associated with the attribute in question for which credit is awarded, for some $k^*<k$. The statistic in (\ref{DVF}) can very simply be amended to take the maximum over the first $k^*$ levels of the attribute.


A potential problem with the current PSM is its dichotomous assessment of stability. That is, if $T(\mathbf{q},\mathbf{P}) \leq \delta$, then the population is deemed stable. However, a small change in the observed value of $\mathbf{P}$ may result in a slight increase in the value of $T(\mathbf{q},\mathbf{P})$ such that $T(\mathbf{q},\mathbf{P}) > \delta$, in which case the population is deemed to be unstable. This problem can be addressed by comparing $T(\mathbf{q},\mathbf{P})$ to various levels of $\delta$, where a smaller value may be require some investigation while a larger value would indicate that the scorecard needs to be redeveloped.

A further potential problem with the use of $T$ is that the measure is more likely to deem an attribute unstable in the case where a large number of small groups are present. That is, a small change in the proportion associated with a small group will more easily be deemed important since the change is rescaled by the proportion $q_j$. This should be taken into account when this measure is used in practice.


\subsection{The Kolmogorov-Smirnov test}\label{KS}

The Kolmogorov-Smirnov test statistic measures the maximum distance between two distribution functions, see \cite{Kol1933}. This well-known test has been studied extensively in the literature and the interested readers is referred to \cite{Con1972} for the case where a the hypothesised distribution is discrete. Note that, although we restrict our attention to the Kolmogorov-Smirnov test here, similar observations are in order when considering other classical goodness-of-fit tests such as the Cram\'er-von Mises test, see \cite{Cra1928} and \cite{VM1928}.

In the current setting, the Kolmogorov-Smirnov test statistics measures the supremum distance between two empirical distribution functions; those corresponding to $\mathbf{q}$ and $\mathbf{P}$. Let $x(j)$ denote the $j^{th}$ level of the attribute under consideration. The empirical distribution function associated with $\mathbf{q}$ is defined to be
\[
    F_{\mathbf{q}}(x(j)) = \sum_{l=1}^j q_l,
\]
where $I(\cdot)$ is the indicator function. Similarly, the empirical distribution function associated with $\mathbf{P}$ is
\[
    F_{\mathbf{P}}(x(j)) = \sum_{l=1}^j P_l.
\]
The Kolmogorov-Smirnov test statistics is defined to be
\[
    D(\mathbf{q},\mathbf{P})=\max_{j \in \{1,\dots,k\}} | F_{\mathbf{q}}(x(j)) - F_{\mathbf{P}}(x(j)) |.
\]

The nature of banking has lead to a situation where large number of customers are scored through the bank's models each month. A typical bank in South Africa can easily score more than 10 000 customers each month for a new product. In the loan loss provisions department of a major financial organization each customer is re-scored through the PD model on a monthly basis, often resulting in millions of new scores each month. This presents a challenge when using standard statistical techniques. The reliance of these techniques on $p$-values proves problematic as a result of phenomena discussed below.

In order to illustrate the problem mentioned above, consider the case where we are interested in testing the hypothesis of population stability for the attribute gender. As a concrete example, consider the case where the a development sample, of size $100 \ 000$, is observed in which $50\%$ of clients are male and $50\%$ are female. At the time of review, another sample of size $100 \ 000$ is observed; but this time $50.5\%$ of the sample are male while $49.5\%$ are female. Using the algorithm outlined in Section \ref{HTBT}, based on $1 \ 000 \ 000$ parametric bootstrap samples, we estimate the $p$-value associated with the Kolmogorov-Smirnov test for population stability to be $0.15\%$. As a result, population stability is rejected at all reasonable nominal significance levels. The rejection of this hypothesis is based on a very small change in the distribution of the attribute in question, and we believe that most practitioners will agree that the observed change should not be considered indicative of an unstable attribute.

It should be noted that classical goodness-of-fit tests are not often used in the assessment of population stability in practice. The authors attribute the lack of popularity of these statistical techniques in the current context to the large sample sizes often encountered in practice. These tests are consistent against fixed alternatives from the null hypothesis in \eqref{eqn:H0}. This means that, if the sample size is large enough, then these tests will reject population stability for any deviation from the null hypothesis, even if this deviation is minuscule.

One final caveat relating to the use of classical goodness-of-fit testing procedures is in order. While the majority of the attributes used in credit scorecards are typically measured on an ordinal scale, there are examples of attributes that are measured on an nominal scale. If this is the case, and the attribute in question has more than two levels, then the use of classical goondness-of-fit techniques is not straightforward. However, since these measures are of limited use in the current context, the required alteration is beyond the scope of the current paper. As a concrete example, consider the application method that a potential customer uses to apply for credit. This attribute may be divided into the following four levels:
\begin{itemize}
    \item Branch - Applications done in the branch.
    \item Online - Application done through an online application channel.
    \item Phone - Applications done through a non-direct channel.
    \item Marketing call - Application done after prompted by the credit provider.
\end{itemize}
When using, for example, the PSI to test population stability, all that is required for the calculation is the observed proportions of each of the levels at the time of development and review of the scorecard. However, computation of $D$ requires the calculation of empirical distribution functions, which inherently assume that the groups are ordered, which is not the case here. This makes the Kolmogorov-Smirnov test difficult to use for the testing of population stability in practice.

As a result of the difficulties mentioned above, we do not include classical goodness-of-fit tests in the numerical examples presented in Section \ref{Numerics}.

\subsection{A measures based on effect sizes}

A possible approach to the problem of testing population stability, which is not affected by the large sample sizes encountered in practice, is based on the use of effect sizes. Under the assumption of population stability; that is, under the assumption that the distribution of an attribute has not changed, we know that $mPs_j \sim Bin(m,p_j)$, for $j \in \{1,\dots,k\}$. As a result, if the population is stable, the expected value and variance of $mP_j$ are $mq_j$ and $mq_j(1-q_j)$, respectively. 

We begin by considering effect sizes in the univariate setting. That is, the effect size of the change between the two populations in the $j^{th}$ level of the attribute is
\[
    \gamma_j = \frac{|P_j-q_j|}{\sqrt{q_j(1-q_j)}}.
\]
Note that this quantity does not depend on $n$ or $m$. That is, given an observed set of values for $\mathbf{q}$ and $\mathbf{P}$, the sample sizes used do not feature in the calculation of $\gamma_j$. When performing this calculation, we obtain a vector of results, $\gamma_1,\dots,\gamma_k$, which we can combine to form a single PSM using a weighted average;
\begin{equation*}
    \Gamma(\mathbf{q},\mathbf{P}) = \sum_{j=1}^k q_j \gamma_j = \sum_{j=1}^k \frac{\sqrt{q_j}|P_j-q_j|}{\sqrt{1-q_j}},
\end{equation*}
where the weight assigned to the effect size associated with the $j^{th}$ level of the attribute is $q_j$; the proportion of that level under the null hypothesis. This weighting is chosen so as to ensure that the contribution of the calculated effect size for level $j$ to $\Gamma$ is proportional to the likelihood of observing this level of the attribute. Since $\Gamma$ is constructed using effect sizes, we can interpret the value of $\Gamma$ along the same lines as an effect size. However, since $\Gamma$ is a weighted average, and we would like this metric to identify changes in the case where the proportions associated with only a small number of levels of the attribute in question change, we recommend using a lower threshold than is used for effect sizes. As a general rule, an effect size is considered to be practically significant if it exceeds $0.3$; for more details, see \cite{Lew1994}. Based on the numerical results presented in Section \ref{Numerics}, we recommend that $\Gamma>0.1$ be considered to indicate a practically significant difference between $\mathbf{q}$ and $\mathbf{P}$.

In order to illustrate the use of this measure and its immunity to the problem of large sample sizes (as experienced by classical goodness-of-fit techniques), we revisit the example included in Section \ref{KS}. If the observed proportions change from $\mathbf{q}=(50\%,50\%)$ to $\mathbf{P}=(50.5\%,49.5\%)$, then the individual effect sizes associated with the two levels are $(\gamma_1,\gamma_2)=(0.01,0.01)$ and $\Gamma(\mathbf{q},\mathbf{P}) = 0.01$. This small value of $\Gamma$ indicates that the difference between $\mathbf{q}$ and $\mathbf{P}$ is not practically significant, meaning that we do not reject the hypothesis of population stability. This is in contrast to the Kolmogorov-Smirnov test for which population stability was rejected.

It should be noted that several different extensions of effect sizes to the multivariate case exist, see \cite{GK2005}. However, we have been unable to find any mention of the use of effect sizes, even in simple settings, as PSMs (with the exception of \cite{DUPISANIEVISAGIE2020} which mentions this possibility without further discussion).


\subsection{Overlapping}

\begin{figure}[!htbp!] \label{fig1}
     \centering
     \includegraphics[width=0.6\textwidth]{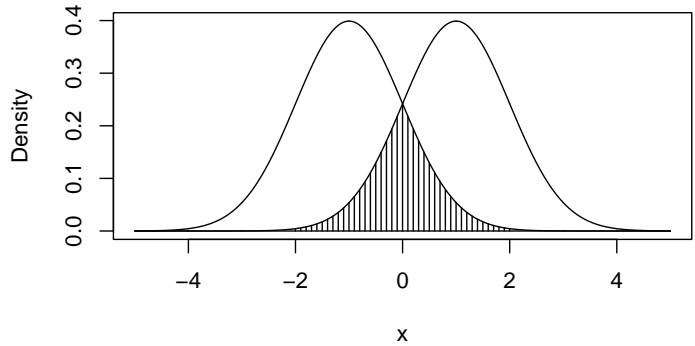}
     \caption{Normal density plots with overlapping area shaded.}
\end{figure}

A different approach for comparing distributions, discussed in \cite{PastoreCalcagni2019}, is a measure of similarity referred to as the overlapping statistic. In the mentioned paper, overlapping, in the case of continuous distributions, is explained to be the area of the intersection of the two corresponding density functions. That is, let $f$ and $g$ be two density functions. The overlapping index between these densities, $\eta(f,g): \mathbb{R} \times \mathbb{R} \rightarrow [0,1]$, is
\[
    \eta(f,g) = \int_{\mathbb{R}} \textrm{min}\{f(x),g(x)\} \textrm{d}x.
\]
This concepts is illustrated using Figure \ref{fig1}. This figure shows two normal density functions, with the ``overlapping'' between these densities shaded.

In the context of population stability, our interest centers on the discrete case; specifically on comparing the mass functions corresponding to $\mathbf{q}$ and $\mathbf{P}$. In this case, the definition of $\eta$ is simply redefined to be
\[
    \eta(\mathbf{q},\mathbf{P}) = \sum_{j=1}^k \textrm{min}\{q_j,P_j\}.
\]

A major advantage of the overlapping statistic is the simple interpretation that it allows; $\eta$ shows the proportion of probability mass that correspond between two distributions. The simplicity of the overlapping concept makes the use of this index particularly suitable for use in the financial sector, given the high level of importance attached to interpretation. The concept of overlapping can easily be explained to a non-technical audience without the need to explain concepts relating to, for example, $p$-values, critical values or effect sizes.
An overlapping index of 0.8 can simple be understood to mean that 80\% of the probability mass between the two distributions under consideration are associated with the same levels of the attribute in question.
Given the emphasis on the interpretability of models in credit scoring and the requirement that the models used be explained to a non-technical audience, we believe that this simple and intuitive measure may prove to be very useful in the context of population stability testing.

\newpage

\section{Numerical results} \label{Numerics}

This section contains three numerical examples which are used to compare the PSMs discussed in Section \ref{PSMs}. The examples considered are as follows:
\begin{enumerate}
    \item A stable population - The distributions of all characteristics remain unchanged.
    \item A stable outcome - The distributions of two of the attributes considered have changed; however, these changes offset each other and the overall PD remains largely unchanged.
    \item An unstable population - The distributions of all attributes included in the scorecard have changed.
\end{enumerate}
It should be noted that all three examples are based on simulated data. The use of simulated data is preferable to the use of observed data in the current context as the former allows us to specify whether or not distributional changes occur. In order to generate the required data sets, we use the simulation technique proposed in \cite{DUPISANIEVISAGIEALLISON2022}.

For each of the examples, we simulate a development population containing ten attributes measured on different scales. Table \ref{Attributes} shows the name, measurement scale and IV for each all of the attributes included in the scorecard. Care is taken to use select attributes often found in a scorecard and to use realistic configurations for the levels of these attributes considered. We use the same attributes and levels as those found in \cite{DUPISANIEVISAGIEALLISON2022}. Since knowledge of the exact distribution of the attributes is not required to interpret the results below (and for the sake of not interrupting the main arguments of the paper), we defer the specification of these distributions to Appendix A. The population size is set at 10 000 customerts throughout.

\begin{table}[!htbp!]%
\caption{The name, measurement scale and information value of the attributes included in the model}
\label{Attributes}
\centering
\small
\begin{tabular}{llc}
\hline
Name & Scale & Information value \\
\hline
Gender & Ordinal & 0.499\\
Existing customer & Ordinal & 0.441\\
Number of enquiries & Ratio  & 0.394\\
Credit cards with other providers & Ratio & 0.515\\
Province of residence & Ordinal & 0.284\\
Application method & Ordinal & 0.222\\
Age & Ratio & 0.164\\
Total amount outstanding & Ratio & 0.083\\
Income & Ratio & 0.182\\
Balance of recent defaults & Ratio & 0.192\\
\hline
\end{tabular}
\end{table}

\subsection{Example 1: A stable population}

In the first example, we simulate a review population from the same distributions as are specified for the development population. The average bad rate for development and review data sets are found to be 9.92\% and 10.08\%, respectively.
Table \ref{tabl:Stable_Pop} shows the calculated results of the various PSMs discussed in Section \ref{PSMs}.


\begin{table}[!htbp!]%
\caption{Stable Population}
\label{tabl:Stable_Pop}
\centering
\small
\begin{tabular}{lcccc}
\hline
Variable & PSI & DPV & Effect Size & Overlapping \\
\hline
Gender & 0.0004 & 0.0241 & 0.0198 & 0.9903\\
Age & 0.0007 & 0.0575 & 0.0102 & 0.9889\\
NumEnq & 0.0010 & 0.0639 & 0.0133 & 0.9874\\
ExistCust & 0.0001 & 0.0153 & 0.0077 & 0.9969\\
CCother & 0.0008 & 0.1206 & 0.0087 & 0.9924\\
OutBal & 0.0008 & 0.1066 & 0.0136 & 0.9878\\
Prov & 0.0016 & 0.1538 & 0.0163 & 0.9838\\
AppMethod & 0.0010 & 0.0458 & 0.0175 & 0.9857\\
Income & 0.0012 & 0.0633 & 0.0128 & 0.9868\\
RecDef & 0.0007 & 0.0734 & 0.0132 & 0.9909\\
\hline 
PD Groups & 0.0017 & 0.0690 & 0.0109 & 0.9837\\
\hline
\end{tabular}
\end{table}

\subsubsection{PSI}

Using the generally accepted rule of thumb for PSI, all variables are considered to be stable, having a PSI lower than 0.1. Using the PSI, we do not reject the hypothesis of population stability.



\subsubsection{DPV}

When using the method proposed in \cite{DUPISANIEVISAGIE2020}, the value of the constant of materiality, $\delta$, is required to be specified. Although it is recommended that this value be chosen by taking business considerations into account, we note that the example presented in \cite{DUPISANIEVISAGIE2020} uses $\delta=0.2$. In order to illustrate the interpretation of this PSM, we also make use of this value. For all attributes considered, we note that the assumption of population stability is not rejected. However, should the value of $\delta$ have been set to $0.15$, the attribute ``Province'' would have been deemed unstable. Note that this attribute has nine levels, some associated with rather small proportions. As was noted above, this combination may conspire to mistakenly suggest distributional changes where none occur. As a result, we caution against using values for $\delta$ smaller than $0.2$ when testing the stability of an attribute with a large number of levels and small associated proportions.






\subsubsection{Effect size}

The PSM based on effect sizes results in small values for each attribute as well as the PD groups. Again, we do not reject the assumption of population stability.

\subsubsection{Overlapping}

The overlapping metric results in values close to $1$ for each of the attributes and the PD groups considered. In fact, the minimum of the observed overlapping statistics observed is in excess of $0.98$. This indicates that the overlapping statistic does not provide evidence agaist the hypothesis of population stability.


\subsection{A stable outcome}

The second set is simulated after changing the distribution of two of the attributes in a way that they would offset one another's effect on the PD. That is, the average PD of the development and review populations are approximately equal. The proportions associated with the levels of ``Number of Enquiries'' as well as ``Credit cards with other providers changed'' are changed in the manner specified in Tables \ref{tab1} and \ref{tab2}, respectively. Table \ref{tabl:Stable_out} shows the outcome of the various PSMs. The table shows that the PSI is calculated to be infinite for the ``Outstanding balance'' attribute. As was pointed out in the discussion in Section \ref{psi}, this is due a level of this attribute being observed in one population but not the other.


\begin{table}[H]%
\caption{Number of enquiries}
\label{tab1}
\centering
\small
\begin{tabular}{clll}
\hline
Group & Description & Development & Review \\
 &  & proportion & proportion \\
\hline
0 & No enquiries & 30\%  & 40\% \\
1 & One enquiry & 25\%  & 25\% \\
2 & Two enquiries & 20\% & 10\% \\
3 & Three enquiries & 15\% & 15\% \\
4 & Four enquiries & 5\% & 5\% \\
5 & Five or more enquiries & 5\% & 5\% \\
\hline
\end{tabular}
\end{table}

\begin{table}[H]%
\caption{Credit cards with other providers}
\label{tab2}
\centering
\small
\begin{tabular}{clll}
\hline
Group & Description & Development & Review \\
 &  & proportion & proportion \\
\hline
0 & No credit cards at another provider & 50\%  & 30\% \\
1 & Credit card at another provider & 30\%  & 50\% \\
2 & Credit cards at another provider & 15\% & 15\% \\
3 & Three or more credit cards at another provider & 5\% & 5\% \\
\hline
\end{tabular}
\end{table}


\begin{table}[!htbp!]%
\caption{Stable Outcome}
\label{tabl:Stable_out}
\centering
\small
\begin{tabular}{lcccc}
\hline
Variable & PSI & DPV & Effect sizes & Overlapping\\
\hline
Gender & 0.0002 & 0.0156 & 0.0128 & 0.9937\\
Age & 0.0014 & 0.0659 & 0.0135 & 0.9848\\
NumEnq & 0.0952 & 0.4948 & 0.1170 & 0.8961\\
ExistCust & 0.0001 & 0.0222 & 0.0112 & 0.9955\\
CCother & 0.2046 & 0.6729 & 0.3323 & 0.7963\\
OutBal & $\infty$ & 0.0410 & 0.0097 & 0.9918\\
Prov & 0.0021 & 0.0775 & 0.0251 & 0.9787\\
AppMethod & 0.0018 & 0.0529 & 0.0269 & 0.9796\\
Income & 0.0021 & 0.2400 & 0.0094 & 0.9878\\
RecDef & 0.0016 & 0.1210 & 0.0250 & 0.9828\\
\hline
PD Groups & 0.0191 & 0.2610 & 0.0372 & 0.9442\\
\hline
\end{tabular}
\end{table}

The average PDs for the development and review populations are 9.92\% and 9.69\%, respectively.
This change can be seen as relatively small and is a result of the distributional changes in the attributes offsetting one another. The changes made to the ``Number of enquiries'' see a greater proportion of the population moving to the lower-risk categories, while the ``Credit cards with other providers'' attribute tends to increase the proportion of the population in the higher risk categories.

\subsubsection{PSI}

The PSI metrics calculated for the two attributes for which the distributions have changed are markedly higher than those calculated for the other attributes. The PSI value associated with the ``Number of enquiries'' attribute is calculated to be 0.0952 while that associated with ``Credit cards with other providers'' has a value of 0.2041. According to the rules-of-thumb described earlier in this paper, the former attribute is considered not to have changed substantially while the latter is considered to have undergone a small distributional change. 

An interesting observation is made for the ``Outstanding balance'' attribute. This attribute contains a bucket for very large balances. In the base set, this bucket has no observations, however in the test set a very small number of observations is seen. The setup of the PSI formula then creates a scenario where if one of the buckets approach zero, the value for PSI will approach infinity. 

\subsubsection{DPV}

The Du Pisanie and Visage metric is successfull in identifying the two attributes whose distributiond have changed. Values of 0.4948 and 0.6729 are calculated for the ``Number of Enquiries and Credit Cards with other providers'' attributes, respectively. As described in \cite{DUPISANIEVISAGIE2020}, the user will need to choose a value for $\delta$ against which to test the values of this metric.  A $\delta$ of 0.5, in this case, would have that the Number of Enquiries are still stable (albeit by a very thin margin) and that the Credit Cards with other providers have moved enough to be deemed unstable. Setting the value of $\delta=0.2$, as was done in the previous example, results in the correct identification of both of the attributes whose distributions have changed. However, the test statistic value for Income is calculated to be $0.24$, meaning that this attribute is deemed to be unstable, even though its distribution has not changed. This can be explained by
the fact that this attribute has a level with a very low associated proportion.

\subsubsection{Effect sizes}

The Effect sizes metric, using a threshold of 0.1 for practical significance, is successful in identifying the attributes that have changed with ``Number of Enquiries'' getting a value of 0.1170 and ``Credit cards with other providers'' getting a value of 0.3323, which is substantially higher than the values calculated for the other metrics. As with the PSI and metric proposed by Du Pisanie and Visagie, the ``Credit cards with other providers'' have a much higher value than the ``Number of enquiries'' for this metric. This corresponds to the greater move seen for the former attribute, compared to the move of the latter attribute.

\subsubsection{Overlapping}

The Overlapping metric clearly shows the exact magnitudes of the changes in the distributions. A value of 0.8961 indicates a move of just over 10\% in the ``Number of Enquiries'' attribute. The ``Credit cards with other providers'' show a value of 0.7963 for the overlapping metric, inline with the 20\% move in the proportions associated with the data.

\subsection{An unstable population}

The final set is simulated using distributions in which six of the attribute sees a 5\% change in the proportions associated with the various buckets, these changes are detailed in Appendix B.
Four attributes (``Gender'', ``Number of Enquiries'', ``Existing Customer'' and ``Credit Cards with other providers'') all see a 5\% shift to higher risk buckets. Two attributes (``Province'' and ``Application method'') see a 5\% move to lower risk buckets. The details pertaining to the move in each attribute can be found in Appendix B. The remaining four attributes (``Age'', ``Outstanding Balance'', ``Income'' and ``Recent defaults'') see no change in the distribution. The average PD has moved from 9.92\% in the baseline data set to 10.67\% in the test data set. The calculated PSMs for this example can be found in Table \ref{tabl:Unstable_out}.


\begin{table}[!htbp!]%
\caption{Unstable Outcome}
\label{tabl:Unstable_out}
\centering
\small
\begin{tabular}{ccccc}
\hline
Variable & PSI & DPV & Effect sizes & Overlapping\\
\hline
Gender & 0.0079 & 0.1092 & 0.0897 & 0.9560\\
Age & 0.0014 & 0.0659 & 0.0135 & 0.9848\\
NumEnq & 0.0175 & 0.1765 & 0.0638 & 0.9461\\
ExistCust & 0.0160 & 0.262 & 0.1321 & 0.9469\\
CCother & 0.0105 & 0.1556 & 0.0738 & 0.9529\\
OutBal & $\infty$ & 0.0410 & 0.0097 & 0.9918\\
Prov & 0.0102 & 0.1462 & 0.0608 & 0.9534\\
AppMethod & 0.0244 & 0.1869 & 0.1001 & 0.9297\\
Income & 0.0021 & 0.2400 & 0.0094 & 0.9878\\
RecDef & 0.0016 & 0.1210 & 0.0250 & 0.9828\\
\hline
PD Groups & 0.0055 & 0.1430 & 0.0197 & 0.9705\\
\hline
\end{tabular}
\end{table}

\subsubsection{PSI}

The PSI metric is able to identify a change in the distribution of all attributes for which the distributions have moved. It is interesting to note that the even though each of the movements in the distributions were similar in magnitude, the calculated values of the PSI differ markedly between the attributes. Consider the attributes ``Province'' and ``Application method''. Both attributes have moved 5\%, however the PSI for Province is 0.0102 and the value for ``Application method'' is 0.0244, a difference factor of close to 2.5 times. Furthermore, it is interesting to note that despite these large differences in the metric values, both attributes will be classified as not having changed substantially, according to the PSI rule of thumb.

\subsubsection{DPV}

The Du Pisanie and Visagie metric is able to correctly identify all the attributes that have moved. As with the PSI metric, some attributes have a much higher value for the calculation than others. The ``Existing customer'' attribute has a value that is 2.5 times higher than the value for ``Gender'', despite the underlying movement being similar. Here we thus notice that the Du Pisanie and Visage metric is prone to similar problems as the PSI in this regard.

The distribution of the ``Income'' attribute has remained unchanged, yet it shows the second highest value for the Du Pisanie and Visagie metric. Closer inspection reveals that one of the groups in the attribute groupings is very small, compared to the other groups, at only 3\% of the population in the base set. In the test set, this group grows by 0.7\% to 3.7\%. Given the small base, this is a substantial growth and results in the variable being deemed very unstable by this metric.


\subsubsection{Effect sizes}

The measure based on effect sizes is successful in identifying the attributes for which the distribution has changed when using a threshold of 0.05 for practical significance. In this example, a cut-off value of 0.1 will result in several of the distributional changes not being identified.

\subsubsection{Overlapping}

The Overlapping metric provides us with the most predictable results and is able to identify the 5\% across most attributes. The attribute with the largest deviation from the expected 0.95 is the Application Method; in this case the value of the statistic is calculated to be 0.9297. 

\section{Summary}


This paper provides an overview of several testing procedures for the hypothesis of population stability. The widely used population stability index (PSI), see \cite{SIDDIQI2006}, is discussed together with some of the problems associated with the use of this index; particular attention is paid to challenges associated with large sample sizes.

In addition to the PSI, newly proposed techniques are considered. These are the measures proposed by Du Pisanie and Visage, see \cite{DUPISANIEVISAGIE2020}, and the population accuracy index proposed by Taplin and Hunt, see \cite{TAPLINHUNT2019}. The former metric exhibits desirable empirical properties; however, it is susceptible to problems associated with very small groupings. The measure proposed by Taplin and Hunt is not well defined for logistic regression and it is shown that the measure can give incorrect answers if specific cases.

The use of Effect sizes is also considered and it is shown that Effect sizes are able to overcome the challenge of large population sizes effectively. Finally, the overlapping statistic is considered. This statistic is very intuitive and simple to use, however, it provides the user with a strong indication of how far the data has shifted.

Of the methods considered for this paper, none provide a way in which to understand the impact of the change on the overall model outcome. All methods are able to ascertain where a single attribute has moved, but cannot do so in relation to the overall outcome. Further research is required in order to develop a metric which measures the magnitude of the change in the predicted outcome as a result of the distributional changes of the individual attributes; the authors are currently in the process of developing such a metric.

\setcounter{section}{5}
\section{Appendix A}

Below we specify the proportions associated with each level of each attribute. For each of these levels, we also give the associated bad ratio. This ratio is used in the determination of the PDs associated with the individual customers using the simulation technique developed in \cite{DUPISANIEVISAGIEALLISON2022}.

\subsection{Gender}

We assume that $60\%$ of applicants are female and $40\%$ are male and we specify the bad ratio of males to females to be $3$.

\subsection{Age}

Table \ref{Age_param} shows the distribution and bad ratios assumed for the attribute, ``age''.
\begin{table}[H]%
\caption{Age}
\label{Age_param}
\centering
\small
\begin{tabular}{lcc}
\hline
Group & Proportion & Bad ratio \\
\hline
18 - 21 & 5\% & 1.00\\
22 - 25 & 7\% & 0.85\\
26 - 30 & 9\% & 0.78\\
31 - 45 & 26\% & 0.66\\
46 - 57 & 21\% & 0.50\\
58 - 63 & 11\% & 0.43\\
64 - 75 & 21\% & 0.31\\
\hline
\end{tabular}
\end{table}

\subsection{Number of enquiries}

Table \ref{NumEnq_param} specifies the proportions and the bad rates for the levels of this attribute.
\begin{table}[H]%
\caption{Number of enquiries}
\label{NumEnq_param}
\centering
\small
\begin{tabular}{clcc}
\hline
Group & Description & Proportion & Bad ratio \\
\hline
0 & No enquiries & 30\%  & 1.0\\
1 & One enquiry & 25\%  & 1.3\\
2 & Two enquiries & 20\% & 1.8\\
3 & Three enquiries & 15\% & 1.9\\
4 & Four enquiries & 5\% & 2.1\\
5 & Five or more enquiries & 5\% & 2.7\\
\hline
\end{tabular}
\end{table}   

\subsection{Existing customer}

$80\%$ of applicants are specified to be existing customers. We assume that the bad ratio for new customers to existing customers is 2.7

\subsection{Credit cards with other providers}

Table \ref{CCOth_param1} shows the assumed distribution of the number of credit cards with other providers, as well as the specified bad ratios.
\begin{table}[H]%
\caption{Credit cards with other providers}
\label{CCOth_param1}
\centering
\small
\begin{tabular}{llcc}
\hline
Group & Description & Proportion & Bad ratio \\
\hline
0 & No credit cards at another provider & 50\%  & 1.0\\
1 & Credit card at another provider & 30\%  & 1.2\\
2 & Credit cards at another provider & 15\% & 1.7\\
3 & Three or more credit cards at another provider & 5\% & 2.5\\
\hline
\end{tabular}
\end{table}  

\subsection{Total amount outstanding}

We assume that the total amount outstanding for a customer follows a standard lognormal distribution, rescaled by a factor of 10 000. The resulting proportions and bad ratios for the levels included in the scorecard can be found in Table \ref{AmtOut_param}.
 \begin{table}[H]%
\caption{Total amount outstanding}
\label{AmtOut_param}
\centering
\small
\begin{tabular}{clcc}
\hline
Group & Grouping & Proportion & Bad Ratio \\
\hline
0 & 0 - 5 000 & 24.4\% & 1.0\\
1 & 5 000 - 10 000 & 25.6\% & 1.2\\
2 & 10 000 - 25 000 & 32.0\% & 2.0\\
3 & 25 000 - 100 000 & 16.9\% & 2.1\\
4 & more than 100 000 & 1.1\% & 0.8\\
\hline
\end{tabular}
\end{table}  

\subsection{Province of residence}

Table \ref{Prov_param} shows the proportions associated with each of the 9 provinces of South Africa, together with the associated bad ratios.
\begin{table}[H]%
\caption{Province of residence}
\label{Prov_param}
\centering
\small
\begin{tabular}{cccc}
\hline
Group & Description & Proportion & Bad ratio \\
\hline
0 & Gauteng & 40\%  & 1.0\\
1 & Western Cape & 30\%  & 0.7\\
2 & KwaZulu Natal & 7\% & 1.8\\
3 & Mpumalanga & 5\% & 1.5\\
4 & North West & 5\% & 3.0\\
5 & Limpopo & 4\% & 2.5\\
6 & Eastern Cape & 4\% & 2.0\\
7 & Northern Cape & 3\% & 4.0\\
8 & Free State & 2\% & 1.2\\
\hline
\end{tabular}
\end{table}   

\subsection{Application method}

Table \ref{AppMeth_param} specifies the distribution and bad ratios of this attribute.
\begin{table}[H]%
\caption{Application method}
\label{AppMeth_param}
\centering
\small
\begin{tabular}{llcc}
\hline
Group & Description & Proportion & Bad ratio \\
\hline
0 & Branch & 30\%  & 1.0\\
1 & Online & 40\%  & 0.5\\
2 & Phone & 15\% & 1.5\\
3 & Marketing Call & 15\% & 0.4\\
\hline
\end{tabular}
\end{table}  

\subsection{Income}

Income is modelled using a mixture distribution with several local models. The associated proportions and bad ratios can be found in Table \ref{Inc_param}.
\begin{table}[H]%
\caption{Income}
\label{Inc_param}
\centering
\small
\begin{tabular}{clcc}
\hline
Group & Grouping & Proportion & Bad Ratio \\
\hline
0 & 0 - 5 000 & 3.2\% & 3.0\\
1 & 5 000 - 11 000 & 15.6\% & 2.5\\
2 & 11 000 - 20 000 & 20.4\% & 2.0\\
3 & 20 000 - 30 000 & 21.8\% & 1.4\\
4 & 30 000 - 70 000 & 24.0\% & 1.2\\
5 & more than 70 000 & 15.0\% & 1.0\\
\hline
\end{tabular}
\end{table}

\subsubsection{Balance of recent defaults}

Table \ref{BalRec_param} specifies the assumed proportions of the values of recent defaults as well as the associated bad ratios.
\begin{table}[H]%
\caption{Balance of recent defaults}
\label{BalRec_param}
\centering
\small
\begin{tabular}{clcc}
\hline
Group & Grouping & Proportion & Bad ratio \\
\hline
0 & 0 - 1 000 & 60.0\% & 1.0\\
1 & 1 000 - 3 000 & 1.1\% & 1.1\\
2 & 3 000 - 5 000 & 2.1\% & 2.0\\
3 & 5 000 - 30 000 & 18.9\% & 2.5\\
4 & 30 000 - 1 000 000 & 18.0\% & 3.0\\
5 & more than 1 000 000 & 0.0\% & 3.3\\
\hline
\end{tabular}
\end{table}

\setcounter{section}{6}
\section{Appendix B}

This appendix contains the details of the changes made to the review population distributions as considered in Section 4.3.

\subsection{Gender}

In the development population there are 60\% females and 40\% males. In the review population there are 55\% females and 35\% males.

\subsection{Number of Enquiries}

Table \ref{NumEnq_unstablechange} specifies the proportions in the development and review populations for the Number of Enquiries attribute.
\begin{table}[H]%
\caption{Number of enquiries}
\label{NumEnq_unstablechange}
\centering
\small
\begin{tabular}{clcc}
\hline
Group & Description & Development proportions & Review proportions \\
\hline
0 & No enquiries & 30\%  & 25\%\\
1 & One enquiry & 25\%  & 30\%\\
2 & Two enquiries & 20\% & 20\%\\
3 & Three enquiries & 15\% & 15\%\\
4 & Four enquiries & 5\% & 5\%\\
5 & Five or more enquiries & 5\% & 5\%\\
\hline
\end{tabular}
\end{table}   

\subsection{Existing Customer}

In the development population 80\% of applicants are specified to existing customers, whilst 20\% of applicants are new customers. In the review population 75\% of applicants are existing customers and 25\% are new customers.

\subsection{Credit cards with other providers}

Table \ref{CCOth_unstablechange} specifies the proportions in the development and review populations for the ``Credit cards with other providers'' attribute.
\begin{table}[H]%
\caption{Credit cards with other providers}
\label{CCOth_unstablechange}
\centering
\small
\begin{tabular}{llcc}
\hline
Group & Description & Development proportions & Review proportions \\
\hline
0 & No credit cards at another provider & 50\%  & 45\%\\
1 & Credit card at another provider & 30\%  & 35\%\\
2 & Credit cards at another provider & 15\% & 15\%\\
3 & Three or more credit cards at another provider & 5\% & 5\%\\
\hline
\end{tabular}
\end{table}  

\subsection{Province of residence}

Table \ref{Prov_unstablechange} specifies the proportions in the development and review populations for the 9 provinces of South Africa.
\begin{table}[H]%
\caption{Province of residence}
\label{Prov_unstablechange}
\centering
\small
\begin{tabular}{cccc}
\hline
Group & Description & Development proportions & Review proportions \\
\hline
0 & Gauteng & 40\%  & 35\%\\
1 & Western Cape & 30\%  & 35\%\\
2 & KwaZulu Natal & 7\% & 7\%\\
3 & Mpumalanga & 5\% & 5\%\\
4 & North West & 5\% & 5\%\\
5 & Limpopo & 4\% & 4\%\\
6 & Eastern Cape & 4\% & 4\%\\
7 & Northern Cape & 3\% & 3\%\\
8 & Free State & 2\% & 2\%\\
\hline
\end{tabular}
\end{table}   

\subsection{Application Method}

Table \ref{AppMeth_unstablechange} specifies the proportions in the development and review populations for the ``Application method'' attribute.
\begin{table}[H]%
\caption{Application method}
\label{AppMeth_unstablechange}
\centering
\small
\begin{tabular}{llcc}
\hline
Group & Description & Development proportions & Review proportions \\
\hline
0 & Branch & 30\%  & 25\%\\
1 & Online & 40\%  & 45\%\\
2 & Phone & 15\% & 15\%\\
3 & Marketing Call & 15\% & 15\%\\
\hline
\end{tabular}
\end{table}  

\bibliographystyle{plain}
\bibliography{lit-AC}

\end{document}